\documentstyle[12pt,epsf]{article}
\newcommand{\be}{\begin{equation}}
\newcommand{\ee}{\end{equation}}
\newcommand{\bea}{\begin{eqnarray}}
\newcommand{\eea}{\end{eqnarray}}
\newcommand{\sirc}[1]{\stackrel{\circ}{#1}}
\makeatletter
\@addtoreset{equation}{section}
\makeatother

\def\appendix#1{
  \setcounter{section}{1}
  \setcounter{equation}{0}
  \renewcommand{\thesection}{\Alph{section}}
  \section*{Appendix \thesection\protect\indent \parbox[t]{11.715cm} {} }
  \addcontentsline{toc}{section}{Appendix \thesection\ \ \ }}

\begin{document}
\begin{flushright}
ITP-SB-99-\\
hep-th/0004123\\
\end{flushright}

\begin{center}
{\LARGE The AdS-CFT correspondence, consistent truncations and 
gauge invariance}
\vskip1truecm

{\large\bf Horatiu Nastase and
 Diana Vaman}\footnote{Research supported by NSF Grant 9722101;  
E-mail addresses: 
hnastase@insti.physics.sunysb.edu and\protect\\
dvaman@insti.physics.sunysb.edu
} \\ {\large
Institute for Theoretical Physics,\\ S.U.N.Y. Stony Brook, NY 11794-3840,USA\\}
\vskip1truecm
\end{center}
\abstract{
\parbox {4.75 in}{We give arguments for a conjecture made in a previous paper,
that one has to use only the gauged sugra action for the calculation of 
correlators of certain operators via the AdS-CFT correspondence. The existence
of consistent truncations implies that the massive modes decouple, and 
gauged supergravity is sufficient for computing n-point functions of CFT 
operators 
coupled to the massless (sugra) sector. The action obtained from the linear 
ansatz, of the type $\phi(x,y)=\phi_I(x)Y^I(y)$ gives only part of the 
gauged sugra. This means that there is a difference for the correlators on the 
boundary of AdS space. We find, studying examples of correlators, that 
the right prescription is to use the full gauged sugra, which implies using 
the full nonlinear KK ansatz. To this purpose,
we analyze 3 point functions of various gauge fields in 5 and 7 dimensions,
and the  
R-current anomaly in the corresponding CFT.
We also show that the 
nonlinear rotation in the tower of scalar fields of Lee et al., 
Corrado et al. and Bastianelli and Zucchini produces a consistent truncation
to the massless level and coincides with the Taylor expansion of the 
nonlinear KK ansatz in massless scalar fluctuations. 
Finally, we speculate about the way to do the full 
nonlinear rotation for the massive tower. 

}}
\newpage
\section{Introduction}

${}$

In two previous papers together with Peter van Nieuwenhuizen \cite{nvv,nvv2},
 we showed that there exists a nonlinear embedding of
7d maximal gauged sugra into 11d sugra and proved the consistency of this 
truncation 11 dimensional fields to the 7 dimensional fields in the $AdS_7
\times S_4$ background. For the $AdS_4\times S_7$ KK reduction of 11d sugra
(to maximal d=4 gauged sugra), de Wit and Nicolai \cite{dwn} proved the 
consistency of the
truncation indirectly (starting from another formulation of 11d sugra, with 
SU(8) invariance). For the $AdS_5\times S_5$ case presumably one can find also 
a consistent truncation of 10d IIB sugra to 5d maximal gauged sugra. 

Based on the existence of these consistent truncations we conjectured in 
\cite{nvv}
that for the computation of correlators via the AdS-CFT correspondence \cite{
mald, gkp, witten},
if we are interested in operators corresponding to gauged sugra fields, it is
enough to take the gauged sugra action. This eliminates an ambiguity in the 
formulation of the correspondence. Let's explain this further: a priori, there
are two ways of dealing with the computation of correlators. The prescription
says to take string theory on the $AdS_p\times S_{D-p}$ background, and 
compute the effective action as a function of the boundary fields. One way 
could  be to take the linear KK expansion in spherical harmonics (given in
\cite{pvn2})
\be
\phi_{Ai}(x,y)=\sum_I\phi_A^I(x)Y^I_i(y)
\ee
and plug it into the 11d sugra action. 

Then the truncation to the subset of fields of interest, $\{ \phi_A^{I_0} \}$
is not consistent in general, because there are terms in the action linear 
in the fields set to zero, $\{ \phi_A^{I_n} \}$. 
That means that their equation of motion, $\delta S/\delta \phi_A^{I_n}=0$,
contains the fields $\phi_A^{I_0}(x)$ as sources, which gives a contradiction.
For the AdS-CFT correspondence, the inconsistency implies that for 
4- and higher-point 
functions of $\phi_A^{I_0}(x)$, all $\phi_A^{I_n}$ will contribute through
Witten diagrams involving the troublesome couplings:

\begin{figure}[hbtp]
\begin{center}
\mbox{\epsfxsize=7cm \epsffile{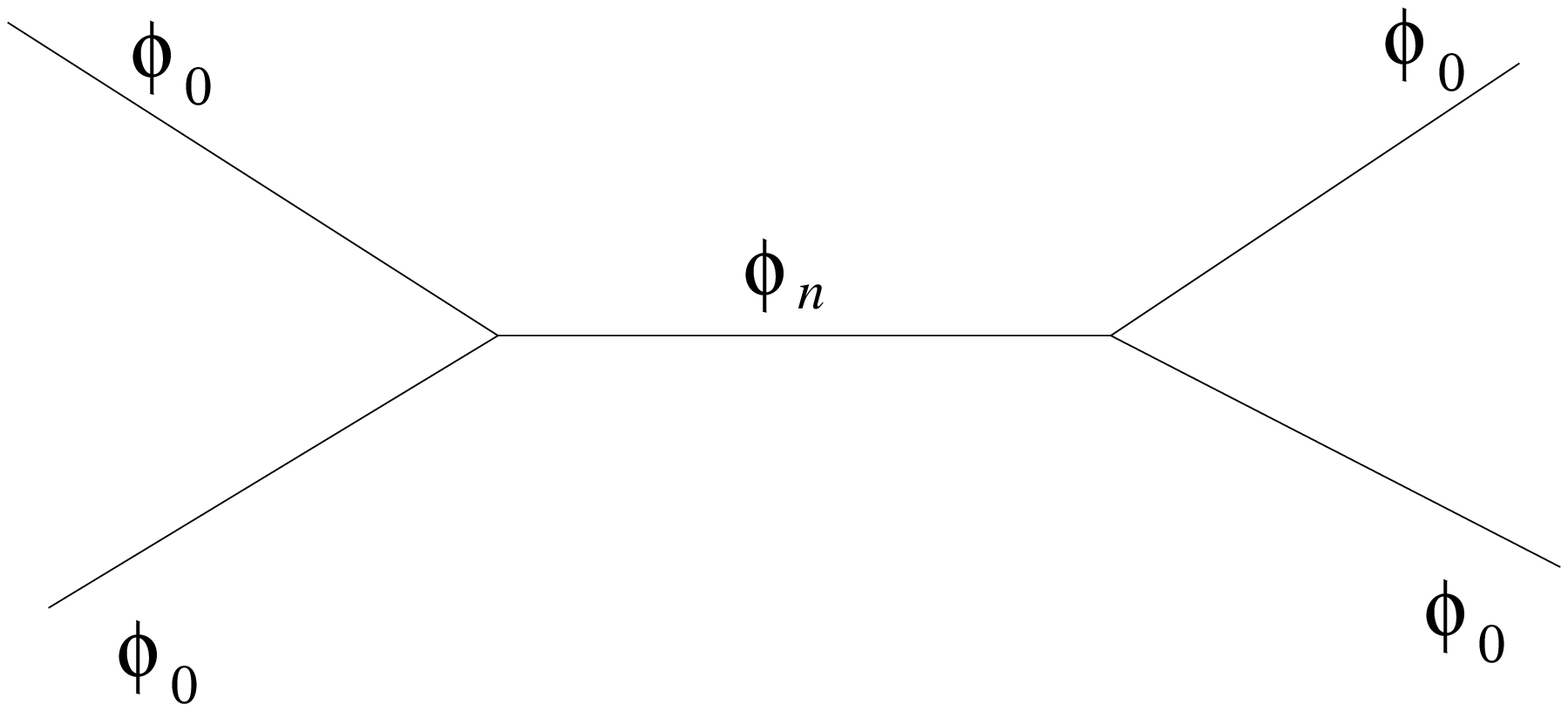}}
\end{center}
\end{figure}

Another possibility appears when we can have a nonlinear ansatz relating the 
$\{\phi_{Ai}(x,y) \}$ to $\{\phi_A^{I_0}(x) \}$ such that the truncation
is consistent (implying in particular that the $\{ \phi^{I_n} \}$ don't
appear in Witten diagrams for $\{\phi^{I_0} \}$).

A priori, we don't know which one to take. We need a physical principle to 
decide. In the cases we study, 11d sugra on $AdS_7\times S_4$  truncated to 
7d gauged sugra and 10d IIB sugra on $AdS_5\times S_5$ truncated to 5d gauged 
sugra, we will argue by examples that it is correct to take the ansatz 
giving a consistent truncation, and not the linear ansatz. In other words,
the gauged sugra action  gives the correct CFT correlators, whereas the 
action coming from the linear ansatz doesn't. At this moment, it becomes clear 
what is the sought-for physical principle. Or rather physical principles: 
gauge symmetry and susy. Indeed, by taking the linearized action for 
the gauged sugra and imposing gauge invariance and susy (by the
Noether procedure)
you obtain the gauged sugra action. In fact, this is how 7d (and 5d) 
gauged sugra were obtained in \cite{ppn,5dgs}.

One might think that taking the gauged sugra action for the calculation of 
correlators is the natural thing to do, but this procedure is available only
if there exists a consistent truncation. If there would exist an inconsistent
truncation to gauged sugra, that would mean that for 4 point correlators 
one would have to consider the contribution of the whole tower of massive 
fields. 

So the procedure one needs in order to obtain the gauged sugra action is to 
modify the linearized ansatz in such a way that the action one obtains is 
gauge invariant and susy. This procedure can be easily generalized. The 
parent action was invariant under local ``gauge'' transformations, with 
parameter $\xi_\mu=\xi^{AB}(x) V_\mu^{AB}(y)$ (where $V_\mu^{AB}$ is a Killing 
vector). After the ``nonlinear redefinition'' (by nonlinear redefinition 
we understand a nonlinear KK ansatz as opposed to a linear one) of the 
massless fields, this invariance is lost, and so we need a corresponding 
nonlinear rotation for the massive fields in order to restore it. It is not 
clear whether this can be done multiplet by multiplet or for the whole tower 
at once. We 
conjecture that this nonlinear ansatz, which we get after performing the 
rotation, is the one needed for the AdS-CFT correspondence.

We have described how to obtain the nonlinear ansatz to be used for the AdS-CFT
conjecture. One possible objection to this procedure is that a nonlinear 
redefinition of fields which doesn't change the quadratic action, like the 
one from the linearized ansatz, $\phi_{Ai}(x,y)=\sum_I\phi_A^I(x)Y^I_i(y)$
to the full nonlinear ansatz, will not change the S matrices of fields. 
This is so in usual field theory, but for the AdS-CFT correspondence there 
is one important difference: the S matrices are for the sources on the 
boundary. And boundary terms, which are usually neglected, become important.
We will show that with a very simple example, of a $\lambda\phi^3$ theory
in the bulk.

Now, to show that taking the gauged sugra is the correct procedure for the 
AdS-CFT correspondence as opposed to taking the action coming from the 
linearized ansatz, we will analyze several n-point functions coming from 
both approaches. We will first analyze in section 2.1 some relevant 3-point 
functions, listing all the possible ones and discussing in particular the 
ones involving gauge fields.  
Then in section 2.2 we will analyze the CS terms and what we can say for 
the field theory anomalies. Finally, we will discuss the scalar 3-point 
functions (corresponding to CPOs) from the work of Lee et al \cite{lmrs}, 
Corrado et al. \cite{cfm}, and Bastianelli and Zucchini \cite{bz} and 
how the nonlinear rotation they found is needed to
obtain a consistent truncation to gauged sugra. We also give arguments on why 
this is just a Taylor expansion in fluctuations of the  full nonlinear 
rotation. 

%%%%%%%%%%%%%%%%%%%%%%%%%%%%%%%%%%%%%%%%%%%%%%%%%%%%%%%%%%%%%%%%%%%%%%%%%%%%%

\section{3-point functions of gauge fields}

\subsection{General considerations}

In this section we will make some general remarks about relevant 3-point 
functions, in particular about gauge fields correlators.

{\bf 7d gauged sugra }

{\em Bosonic fields}: gauge fields $B_{\alpha}^{AB}$ with gauge group 
$SO(5)_g$, antisymmetric tensors
$S_{\alpha\beta\gamma ,A}$, graviton $e_{\alpha}^I$, scalars ${\Pi_A}^i$ in the
coset $Sl(5,{\bf R})/SO(5)_c$.

{\em Bosonic action}:

\bea
&&e^{-1}{\cal L}=-\frac{1}{2}R+\frac{1}{4}m^2(T^2-2T_{ij}T^{ij})-\frac{1}{2}
P_{\alpha ij}P^{\alpha ij}-\frac{1}{4}({\Pi_A}^i{\Pi_B}^j 
F_{\alpha\beta}^{AB})^2
\nonumber\\
&+&\frac{1}{2}({{\Pi^{-1}}_i}^AS_{\alpha\beta\gamma ,A})^2 
+\frac{1}{48}me^{-1}\epsilon^{\alpha\beta\gamma\delta\epsilon\eta\zeta}
\delta^{AB}S_{\alpha\beta\gamma ,A}F_{\delta\epsilon\eta\zeta ,B}\nonumber\\
&+&\frac{ie^{-1}}{16\sqrt{3}}\epsilon^{\alpha\beta\gamma\delta
\epsilon\eta\zeta}
\epsilon_{ABCDE}\delta^{AG}S_{\alpha\beta\gamma, G}F_{\delta\epsilon}^{BC}
F_{\eta\zeta}^{DE}
\nonumber\\
&+&\frac{m^{-1}}{8}e^{-1}\Omega_5[B]-
\frac{m^{-1}}{16}e^{-1}\Omega_3[B]\label{7action}
\eea

The first remark is that gravity will appear in the correct way just because
of general coordinate invariance, both in the linearized ansatz and in the 
nonlinear one. (or rather, the 11d graviton will be nonlinearly redefined 
-- Weyl rescaled --  but only by the scalars: 
$e_{\alpha}^a\rightarrow e_{\alpha}^a [det
E_{\mu}^m]^{-1/5}$). So we will disregard the 3-point functions involving the 
graviton. Also, the $(\delta{\Pi_A}^i)^3$ 3-point function will be analyzed 
in the last section. The remaining 3-point functions are:

-Involving scalars: $B_{\alpha}^{AB}\delta {\Pi_A}^i\partial_{\alpha}\delta
{\Pi_B}^i$, from $[P_{\alpha ij}]^2$ term, $BB\delta\Pi$, from the $dB\wedge
*dB\delta\Pi$ piece in $({\Pi_A}^i{\Pi_B}^i F_{\alpha\beta}^{AB})^2$ and 
$SS\delta \Pi  $, from the $[{(\Pi^{-1})_i}^A S_{\alpha\beta\gamma ,A}]^2$
term.

-Involving no scalars: $(B_{\alpha}^{AB})^3$, from the kinetic term, 
$2*dB\wedge B\wedge B$; SSB, from the S kinetic term, i.e. $
\frac{1}{48}me^{-1}\epsilon^{\alpha\beta\gamma\delta\epsilon\eta\zeta}
\delta^{AB}S_{\alpha\beta\gamma ,A}F_{\delta\epsilon\eta\zeta ,B}$, and BBS, 
from the term $
\frac{ie^{-1}}{16\sqrt{3}}\epsilon^{\alpha\beta\gamma\delta
\epsilon\eta\zeta}
\epsilon_{ABCDE}\delta^{AG}S_{\alpha\beta\gamma, G}F_{\delta\epsilon}^{BC}
F_{\eta\zeta}^{DE}$.

Let's look at the $B^3$ term. The calculation of $AdS$ space correlators
of gauge fields was done in \cite{fmmr,cnss}.

Let's see what would happen if we took the linearized ansatz: 

For the $B^3$ term the $AdS$ space, the calculation of correlators was done in 
\cite{fmmr,cnss}. In 7d, the $*dB
\wedge B\wedge B$ term comes in the nonlinear ansatz in part 
from the kinetic
term $F_{\alpha\beta\mu\nu}^2$ in d=11, and so this piece
 will be absent if we take the linearized ansatz. But there is also a piece 
coming from $\int \sqrt{G^{(11)}}R^{(11)}$, which will remain. So the 
coefficient of the CFT correlator of 3 R-currents would get modified. Although
the CFT has no lagrangean formulation, one can think of making a free field 
calculation, as it was done for the correlators of stress tensors in 
\cite{bft}.
The coefficient would not be fixed, but it can be fixed by taking susy 
variations on the stress tensor correlator in \cite{bft}. One should obtain
the result matching the AdS 3-point function in \cite{fmmr,cnss}.

So the correct result is the one coming from  the nonlinear ansatz. 
Moreover, we clearly see that
imposing gauge invariance on $dB\wedge *dB$ we get the usual $dF\wedge *dF$ 
action, so gauge invariance here is clearly the  physical principle
needed to modify the linearized ansatz. 

The same comment applies to the $BBS$ correlator: The $\epsilon SFF $ term in 
the action comes from two sources: the $\epsilon^{\mu\nu\rho\sigma}
\epsilon^{\alpha_1...\alpha_7}{\cal A}_{\alpha_1\alpha_2\alpha_3}
F_{\mu\nu\rho\sigma}F_{\alpha_4...\alpha_7}$ and 
 $\epsilon^{\mu\nu\rho\sigma}
\epsilon^{\alpha_1...\alpha_7}{\cal A}_{\alpha_1\alpha_2\alpha_3}\\
F_{\mu\nu\alpha_4\alpha_5}F_{\rho\sigma\alpha_6\alpha_7}$. If we use the 
linear ansatz, the last term would give the correct piece, $\epsilon S\partial
B \partial B$, but the former would not contribute, and so the normalization
of the $S \partial B\partial B$ correlator would be wrong. Here one would have 
to compute the $AdS$ correlator first, which we leave for future work 
\cite{nv4}.

{\bf 5d gauged sugra}

{\em Bosonic fields}: -ungauged model: gravitons $e_{\mu}^r$, gauge fields 
$A_{\mu}^{AB}$ scalars $V_{AB}\;^{ab}$ (27-bein), 
global symmetry group $E_{6(6)}$,
composite symmetry USp(8).

                      -gauged model: $e_{\mu}^r$, gauge fields $A_{\mu}^{IJ},
B_{\mu}^{I\alpha}$, scalars $V^{IJab},V_{I\alpha}\;^{ab}$, gauge group:
$SO(p,6-p)\times SL(2,R)$, composite symmetry USp(8). Under $27\rightarrow
(15,1)\oplus (6,12), AB\rightarrow IJ\oplus I\alpha$.

{\em Bosonic action}:
\bea
e^{-1}{\cal L}_{bosonic}&=&-\frac{1}{4}R+\frac{1}{24}P_{\mu abcd}P^{\mu abcd}
\nonumber\\&&-\frac{1}{8}(F_{\mu\nu ab}+B_{\mu\nu ab})^2
-\frac{1}{96}\epsilon^{\mu\nu\rho\sigma\tau}\eta_{IJ}\epsilon_{\alpha\beta}
B_{\mu\nu}^{I\alpha}D_{\rho}B_{\sigma\tau}^{J\beta}\nonumber\\
&&+g^2[\frac{6}{45^2}T_{ab}^2-\frac{1}{96}A_{abcd}^2]+\frac{1}{12}
\epsilon^{\mu\nu\rho\sigma\tau}\epsilon^{IJKLMN}\nonumber\\&&
(F_{IJ\mu\nu}F_{KL\rho\sigma}A_{MN\tau}+g\eta^{PQ}F_{IJ\mu\nu}A_{KL\rho}A_
{MP\sigma}A_{QN\tau}\nonumber\\
&&+\frac{2}{5}g^2\eta^{PQ}\eta^{RS}A_{IJ\mu}A_{KP\nu}A_{MR\sigma}A_{SN\tau})
\eea
where in the ungauged model $\tilde{V}_{cd}\;^{AB}\partial_{\mu}V_{AB}\;^{ab}
=2Q_{\mu[c}\;^{[a}\delta_{d]}^{b]} +P_{\mu}\;^{ab}\;_{cd}$, which becomes in 
the gauged model $\tilde{V}D_{\mu}'V=2Q_{\mu}+P_{\mu}$, and $\tilde{V}=
V^{-1}$, $D_{\mu}'$ is the SO(p, 6-p) covariant derivative, $Q_{\mu}$ is the
USp(8) connection, $D_{\mu}$ is the full $USp(8)\times SO(p,6-p)$- covariant
connection. Also,
\bea
A_{abcd}&\equiv& T_{a[bcd]},\;\; T_{ab}=T^c\;_{abc} \nonumber\\
T^a\;_{bcd}&\equiv&Y^{ae}\;_{becd}=(2V^{IKac}\tilde{V}_{beJK}-V_{JK}\;^{ac}
V_{be}\;^{I\alpha})\eta^{JL}\tilde{V}_{cd IL}\nonumber\\
F_{\mu\nu}^{ab}&=&V^{IJab}F_{\mu\nu IJ},\;\; B_{\mu\nu}\;^{ab}=V_{I\alpha}\;
^{ab}B_{\mu\nu}^{I\alpha}
\eea
and $F_{\mu\nu IJ}$ and $B_{\mu\nu}^{I\alpha}$ are the field strengths of 
$A_{\mu}^{IJ}$ and $B_{\mu}^{I\alpha}$, respectively.

{\em Bosonic 3-point functions}- except the ones involving  the graviton, 
for the same reasons as in 7d, and the ones involving only scalars, which are 
treated in the last section. 

-Involving no scalars: $AAA$, from the $*dA\wedge A\wedge A$ and $dA\wedge 
dA\wedge A$ terms in the action, $BBA$ from the $*dA\wedge B\wedge B$ and $
dB\wedge dB\wedge A$ term in the action (namely from the $\epsilon BDB$ 
term), $BBB$ from the $*dB\wedge B\wedge B $
term, and $BAA$ from the $*dB\wedge A\wedge A$term.

-Involving scalars: $VVA$ terms from the $\partial V\partial V A$ piece of $
P_{\mu ij}^2 $ ($A_{\mu}^{IJ}$ coming from $D'_{\mu}$), and $BBV$ from $
\frac{1}{96}\epsilon^{\mu\nu\rho\sigma\tau}\eta_{IJ}\epsilon_{\alpha\beta}
B_{\mu\nu}^{I\alpha}D_{\rho}B_{\sigma\tau}^{J\beta}$ (V coming from the 
$Q_{\mu}$ term in $D_{\lambda}$).

The $AAA$ term was computed in \cite{fmmr,cnss} 
and gives the correct CFT correlators (we should stress once again that the 
agreement between the AdS and CFT computations holds as long as one uses the 
gauged sugra interactions). We can easily extend this result to 
all the 3 point functions of gauge fields ($BBA$, $BAA$ and $BBB$), 
and all that changes 
are the coefficients of the terms in the action involving gauge fields, and 
the combinatorial factors (coming from differentiating with respect 
to the boundary sources of the gauge fields).

On the other hand, if we take the 10d IIB sugra action,
\bea
S_{IIB, 10d}&=&\frac{1}{(2\pi)^7\alpha '\;^4}\int d^{10}x\{\sqrt{-g^{(10)}}e^{-
2\phi}({\cal R}^{(10)}+4|d\phi|^2-\frac{1}{3} |H|^2)-2|dl|^2-\nonumber\\&&
\frac{1}{3} |H'-lH|^2-\frac{1}{60} |M^+|^2]-\frac{1}{48} C^+\wedge H\wedge H'\}
\label{10dac}
\eea
and plug in the linearized ansatz,
we will again miss some terms of the type $*dA\wedge A\wedge A$ coming from 
the kinetic term$|H'-lH|^2$ of the antisymmetric tensors. The same comment 
applies to the $*dB\wedge B\wedge B$ term for the $BBB$ correlator, the 
$*dB\wedge A\wedge A$ term for the $BAA$ correlator, and $*dA\wedge B\wedge B$
for the $ABB$ correlator. The fact that we don't know the nonlinear KK 
embedding it's not relevant, because we know that if we have a consistent
truncation, the prescription we suggest is to use the gauged sugra action.
We also know the linearized KK reduction of \cite{krn}. So we can say that the 
correlators obtained from the linearized ansatz will differ from the 
ones obtained 
from the nonlinear ansatz, which we know to be correct (i.e. in agreement 
with $N=4$ SYM results). Once again, the 
nonlinear ansatz is seen to be the correct one to take.

%%%%%%%%%%%%%%%%%%%%%%%%%%%%%%%%%%%%%%%%%%%%%%%%%%%%%%%%%%%%%%%%%%%%%%%%%%%%%%

\subsection{Anomalies}

{\em 5 dimensions} For the relation between the CS term in maximal 5d 
gauged sugra and the R-current anomaly in 4d N=4 SYM, Witten gave a very 
elegant argument in his original paper on the AdS-CFT correspondence 
\cite{witten}. The 
argument goes as follows:

If we vary the bulk gauge fields (in 5d) by $\delta_{\Lambda} A_{\mu}^a(x)
=(D_{\mu}\Lambda)^a(x)$ the only nonzero term in the variation of the action
$\delta_{\Lambda}S_{cl}[A_{\mu}^a(x)[A_i^a{\vec{x}}]]$ will be a boundary 
term coming from the CS term,

\bea
\delta_{\Lambda}S_{cl}&=&\delta_{\Lambda}S_{CS}=\int d^4x \Lambda^a(\vec{x})
(\frac{-ik}{96\pi^2})d^{abc}\epsilon^{ijkl}\partial_i(A_j^a\partial_k A_l^c
\nonumber\\&&
+\frac{1}{4}f^{cde}A_j^bA_k^dA_l^e)
\eea

And the conjecture implies that $S_{cl}[A_{\mu}^a(x)] $ is equal to $W[A_i^a
(\vec{x})]$, the generating functional of connected Green's functions   on 
the boundary. Since also $J_i^a(\vec{x})=\delta W[A]/\delta A_i^a(\vec{x})$,
we get 
\be
\delta_{\Lambda}S_{cl}=\delta_{\Lambda}W=-\int d^4x \Lambda^a(\vec{x})D_iJ_i
(\vec{x})
\ee
which implies that 
\be
<D_iJ_i(\vec{x})>=\delta_{\Lambda}S_{CS}
\ee
and gives a concrete physical interpretation to the known mathematical 
fact that the consistent anomaly in n dimensions is obtained by a descent 
equation from the 2n+1 dimensional CS action. 

That implies that if one takes the full 1-loop 3-point function of R-currents
in SYM and one takes a divergence, it should reproduce the 
result for the 'Witten
diagram' of 3 gauge fields in AdS, with a divergence taken. (That is because 
the anomaly is only 1 loop, by the Adler-Bardeen theorem.) It is indeed so, as
noted in
\cite{fmmr,cnss}; but it also implies a similar result for 
the 4 point function.
In 4d N=4 SYM, the box diagram and its anomalous part are also nonzero. 
A priori, there seems to 
be another diagram contributing an anomaly, but it actually gives only a 
renormalization. 
\begin{figure}[hbtp]
\begin{center}
\mbox{\epsfxsize=10cm \epsffile{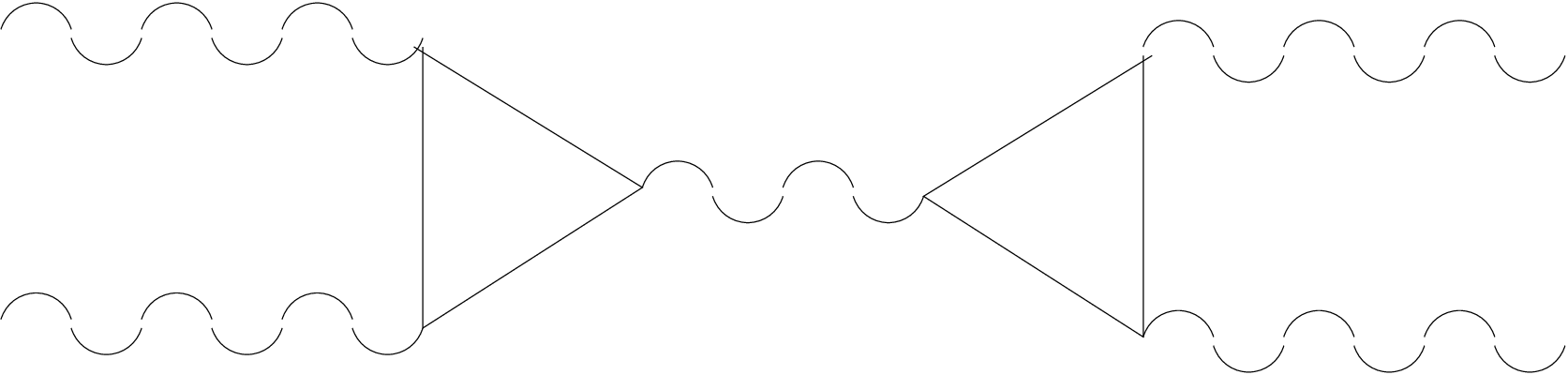}}
\end{center}
\end{figure}
(The wavy lines denote gauge vector propagators while the straight lines 
indicate fermion propagators.)
 
Only by summing all the one loop anomalous diagrams we get a gauge
covariant anomaly \cite{asmo} 
\be
(D^\mu J_\mu)^a=\frac{N^2-1}{384\pi^2}i d^{abc}\epsilon^{\mu\nu\rho\sigma}
F_{\mu\nu}^b F_{\rho\sigma}^c
\ee
(the triangle anomaly anomaly alone yields just the  $dA\wedge dA$ term on 
the right hand side).

On the AdS side, there are two diagrams, the 4 point vertex for the CS term 
and the exchange diagrams, with 3-point vertices coming also from the CS term.
The naive expectation,  that the AdS 4-point vertex equals the box diagram 
from field theory, and the AdS exchange diagram equals the diagram with 
two triangles glued, is wrong, because the AdS exchange diagram gives a 
genuine contribution, not just a renormalization.

However, Witten's argument tells us that the sum of the AdS diagrams 
should be equal to the sum of the field theory diagrams, so we don't need 
to worry. 

What would happen if we take the linearized ansatz instead? Even if we 
don't know the nonlinear ansatz, we know that using the nonlinear ansatz, the
action for the massless sector will be 5d maximal gauged sugra, which has
a CS term, and therefore generates a R-current anomaly. 
On the other hand, using the linear ansatz \cite{krn}, and substituting
it in the 10d IIB sugra action, we notice that there is no surviving CS term! 
That is so because the linear ansatz of the 3-form field strengths $H$ and 
$H'$ does not contain the massless gauge vector fields.
So, by using the linear ansatz, one would conclude that there is no 
R-current anomaly.
 
And again, gauge and supersymmetry invariance tells us that we should take the 
nonlinear ansatz, because by imposing them both on the linearized action we
get the 5d maximal gauged sugra, with a CS term. 

{\em 7 dimensions} Witten's argument applies equally well in all odd 
dimensions, relating the anomaly in 2n dimensions to a CS term in a gauged 
sugra in 2n+1 dimensions. However, the only other example of maximal 
gauged sugra in 2n+1 dimensions is d=7. Again, by varying

\be
S_{CS}=\frac{m^{-1}}{8}e^{-1}\Omega_5[B]-
\frac{m^{-1}}{16}e^{-1}\Omega_3[B]
\ee
where $\Omega_3[B]$ and  $\Omega_5[B]$ are the Chern-Simons forms for 
$B_{\alpha}^{AB}$ (normalized to $d\Omega_3[B]=(TrF^2)^2$ and 
$d\Omega_5[B]=(TrF^4)$), we should get the chiral anomaly in 6d. 
But now it is unclear how to compute
this anomaly, since the dual 6d theory is a 
nontrivial (0,2) CFT without a lagrangean formulation. The anomaly means that
the $SO(5)$ R-symmetry, which is part of the susy algebra, is broken by the
fact that correlators of R-currents are anomalous. In 6d, the first anomalous
correlator is the 4-point function (corresponding to the 4-point CS coupling
$dB\wedge dB\wedge dB\wedge B$). And it is easy to see from the nonlinear 
ansatz \cite{nvv,nvv2} that the 4-point CS coupling (in fact, all the 
CS term!) will be missing for the linearized ansatz (the 7 dimensional 
CS has terms with at least four fields, $Tr(dB\wedge dB\wedge dB\wedge B)$ 
and $Tr(dB\wedge dB)\wedge Tr(dB\wedge B)$, while the 11 dimensional CS
$dF^{(4)}\wedge dF^{(4)}\wedge A^{(3)}$ cannot generate them after substituting
a linearized ansatz).

But we know that the 6d (0,2) CFT should have a chiral anomaly, because it is 
obtained as the IR limit of the M5-brane theory, which has an anomaly.
Moreover, in \cite{hmm}, a brane calculation of the anomaly was performed,
and it was found that the anomaly has the expected functional dependence, i.e.
coming by descent formalism from the 7d Chern-Simons term. The only nontrivial
aspect is the coefficient in front of this anomaly, which was found to be
proportional to $N^3$. The same $N^3$ dependence is found also from 
the AdS calculation, because the sugra coupling has this dependence. One would 
like to understand the $N^3$ dependence in a field theory context, because
the calculation in \cite{hmm} uses M theory, as does the AdS-CFT calculation.
However, that was not done yet. The free-field calculation in \cite{bft} for
the stress-tensor correlators (trying to match with the anomaly calculation
of \cite{hs} on the AdS side), and the free-field calculation in \cite{mape}
for the R-current anomaly, both impose by hand the $N^3$ dependence. In
\cite{intrili}, the \cite{hmm} calculation was extended to other gauge groups,
and in \cite{bb} the calculation was related to Witten's \cite{witten2}
calculation in type IIA string theory, but a real field theory explanation 
is still lacking.

So again, since we want an anomaly in 7d, because we know it should be there 
by the AdS-CFT correspondence, 
the nonlinear ansatz is the correct one. As before, we need to impose 
both susy and gauge invariance on the linearized action obtained by 
compactification in order to recover the correct result. (The absence of a 
CS term respects gauge invariance alone.)

%%%%%%%%%%%%%%%%%%%%%%%%%%%%%%%%%%%%%%%%%%%%%%%%%%%%%%%%%%%%%%%%%%%%%%%%%%%%%%
\section{Scalar 3-point functions}

Let's start by giving the $\lambda\phi^3$ example as promised in the 
introduction. \footnote{This example emerged in a discussion we had with 
Fiorenzo Bastianelli.} For a $\lambda\phi^3$ theory in the bulk, a 
redefinition 
of fields in the bulk doesn't change the bulk S matrices, but does the ones 
computed on the boundary (via the AdS-CFT-type correspondence), because of the
presence of a boundary term.  Let's start with the lagrangean

\be
{\cal L}= \frac{1}{2}(\partial_{\mu}\phi)^2+\frac{1}{2}m^2\phi^2+\lambda
\phi^3\label{toy}
\ee
The bulk 3-point function will be equal to $\lambda$. Let's now redefine 
$\phi=\tilde{\phi}+a\tilde{\phi}^2$. Then,
\bea
{\cal L}&=&\frac{1}{2}(\partial_{\mu}\tilde{\phi})^2+\frac{1}{2}m^2\tilde
{\phi}^2+\lambda\tilde{\phi}^3 +2a\tilde{\phi}(\partial_{\mu}\tilde{\phi})^2
+m^2a\tilde{\phi}^3\nonumber\\
&&+2a^2\tilde{\phi}^2(\partial_{\mu}\tilde{\phi})^2+\frac{1}{2}m^2\tilde{\phi}
^4+3\lambda a\tilde{\phi}^4\nonumber\\
&&+3\lambda a^2\tilde{\phi}^5+\lambda a^3\tilde{\phi}^6\label{toy2}
\eea
But
\be
\int[2a\tilde{\phi}(\partial_{\mu}\tilde{\phi})^2+m^2a\tilde{\phi}^3]
=-a\int \tilde{\phi}^2\Box\tilde{\phi}+\int m^2a\tilde{\phi}^3 
+2a\int\partial_{\mu}(\tilde{\phi}^2\partial_{\mu}\tilde{\phi})\label{boundary}
\ee
On shell, $(\Box-m^2)\tilde{\phi}=0$, so on-shell, the 3-point 
correlator is obtained from 
\be
\int\lambda\tilde{\phi}^3+a\int\tilde{\phi}^2(\Box-m^2)\tilde{\phi}=
\lambda\int\tilde{\phi}^3
\ee
and therefore the correlator is still equal to $\lambda$ 
(For the 4-point correlator,
the calculation is a bit more involved, but the result is the same.).
 But we 
see that if we compute $S_3[\tilde{\phi}|_{bd}]$ to get the 3-point 
correlator of the boundary theory, it will differ from $S_3[\phi|_{bd}]$ by
$\int\partial_{\mu}(\tilde{\phi}^2\partial_{\mu}\tilde{\phi})$. So a 
nonlinear redefinition of bulk fields, which doesn't change the masses, does 
change the boundary correlators.

However, the following observation was made in \cite{dfmmr}. The extra term 
\be
\int_{M}\partial_{\mu}(\phi^2\partial_{\mu}\phi)=\int_{\partial M}\phi (x)
\phi (x)\partial _{\mu }\phi (x) 
\ee
will contribute only a contact term to correlators, because we have
\be
\frac{\delta S}{\delta \phi (x)\delta\phi (y)\delta\phi (z)}\propto
\delta (x-y)
\ee
where $\phi (x)$ here lives on the boundary. Still, if we have the correct 
action from the start, no unwanted contact terms will appear. But we had 
only contact terms due to the simplicity of the example. The field 
redefinition used in \cite{lmrs,cfm} involves also derivatives, and is of 
the type $\phi=\tilde{\phi}+a\tilde{\phi}^2 +b(\partial_{\mu}\tilde{\phi})
^2$. Then, substituting into (\ref{toy}), we get an extra cubic term to be 
added to (\ref{toy2}) (and some higher order terms too)
\be
\int [2b\partial_{\mu}\tilde{\phi}\partial^{\mu}\partial^{\nu}\tilde{\phi}
\partial_{\nu}\tilde{\phi}+ m^2b \tilde{\phi}(\partial_{\mu}\tilde{\phi})^2
\ee
which can be rewritten by partial integration as
\be
b\int(\partial_{\mu}\tilde{\phi})^2(m^2-\Box)\tilde{\phi}
-b\int \partial_{\nu}((\partial_{\mu}\tilde{\phi})^2\partial_{\nu}\tilde{\phi})
\label{boundary2}
\ee
so again, on shell we get only a boundary term contribution to the 3-point
correlator. But this time it is not just a contact term, as it was also noticed
in \cite{dfmmr}. We will come back to the discussion of \cite{dfmmr}
at the end of this section.

Let's now turn to the 3-point functions of scalars. 

{\em 7 dimensions} The gauged sugra scalar fields in 7 dimensions are 
described by a coset element ${\Pi_A}^i \in SL(5,{\bf R})/SO(5)_c$. In the 
physical gauge, it is symmetric and traceless. In terms of scalar fluctuations,
$\delta \pi_{Ai}$, we can write:
\bea
{\Pi_A}^i&=&e^{\delta\pi_{Ai}}=[1+\delta\pi+\frac{\delta\pi^2}{2}+
\frac{\delta\pi^3}{3!}+...]_{Ai}\nonumber\\
{(\Pi^{-1})_i}^A&=&e^{-\delta\pi_{Ai}}=[1-\delta\pi+\frac{\delta\pi^2}{2}-
\frac{\delta\pi^3}{3!}+...]_{Ai}
\eea
And so 
\bea
T_{ij}&=&[1-2\delta\pi+2\delta\pi^2-\frac{4}{3}\delta\pi^3+\dots]_{ij}
\nonumber\\
TrT&=&5+2Tr\delta\pi^2-\frac{4}{3}Tr\delta\pi^3+\dots
\eea
where $T_{ij}={(\Pi^{-1})_i}^A{(\Pi^{-1})_j}^A$. From $P_{\alpha ij}=
[\Pi^{-1}\partial_{\alpha}\Pi ]_{(ij)}$, we get by expansion
\be
P_{\alpha ij}P^{\alpha ij}=Tr(\partial_{\alpha}\delta\pi)^2+0-
\frac{5}{6}Tr[(\Box\delta\pi)\delta\pi^3]-\frac{1}{2}Tr[(\partial_{\alpha}
\delta\pi)^2\delta\pi^3]
\ee
We notice that the cubic terms in $P_{\alpha ij}P^{\alpha ij}$ cancel, but 
the quartic ones don't. 

So, the cubic action for the scalars in the 7d gauged sugra is 
\be
-\frac{1}{2}P_{\alpha ij}P^{\alpha ij}+\frac{1}{4}(T^2-2T_{ij}T^{ij})
=-\frac{1}{2}Tr(\partial_{\alpha}\delta \pi)^2+Tr\delta\pi^2+2Tr\delta\pi^3
\ee

Let us  describe 
the work Corrado et al. \cite{cfm} and Bastianelli 
et al. \cite{bz} in 7 dimensions for computing correlators of 
CPOs in the boundary CFT from the scalar fields  correlators in AdS space 
and see what one can learn from this (This procedure was introduced for 
the first time by Lee et al. \cite{lmrs} for the study of $AdS_5$/N=4 SYM 
correspondence.).

Again, if one compactifies the 11d sugra action on $AdS_7\times S_4$ as in 
\cite{pvn2, bz} , one can write an ansatz for the fields as 
\bea
G_{\Lambda\Pi}&=&\sirc g_{\Lambda\Pi}+h_{\Lambda\Pi}\nonumber\\
h_{\alpha\beta}&=&h'_{(\alpha \beta )}+(\frac{h'}{7}-\frac{h_2}{5})\nonumber\\
h_{\mu\nu}&=&h_{(\mu\nu)}+\frac{h_2}{4}g_{\mu\nu}
\nonumber\\
F^{(4)}&=&\sirc{F^{(4)}} +d a^{(3)}
\eea
where the notation $\sirc g_{\Lambda\Pi}$ means background metric and 
$h_{(\mu\nu)}$ 
stands for symmetric and traceless. The indices $\Lambda ,\dots$ are 11d, 
$\alpha ,\dots$ are $AdS_7$ and $\mu ,\dots$ are $S_4$ indices. One 
decomposes the sugra fields (in the gauge $\sirc D^\mu 
h_{(\mu\nu)}=\sirc D^\mu h_{\mu\alpha}=\sirc D^\mu a_{\mu\Lambda\Pi}=0)$ 
in spherical harmonics:
\bea
h'_{\alpha\beta}&=&\sum h'_{(\alpha\beta) I}Y^I\nonumber\\
h_{(\mu\nu)}&=&\sum \phi^IY^I_{(\mu\nu)}\nonumber\\
h'&=&\sum h'^IY^I\nonumber\\
h_2&=&\sum h_{2I}Y^I\nonumber\\
a_{\mu\nu\rho}&=&\sum b^I\epsilon_{\mu\nu\rho\sigma}D^{\sigma}Y_I
\eea
where $\Box_x Y^I(x)=-k(k+3)Y^I(x)$.
The scalar kinetic term is diagonalized by the eigenvectors
\bea
s^I&=&\frac{k_1}{2k_1+3}(h_2^I+32/\sqrt{2}(k+3)b^I)\nonumber\\
t^I&=&\frac{k+3}{2k+3}(h_2^I-32/\sqrt{2}kb^I)\nonumber\\
\Box s^I&=&k(k-3), k\ge 2\nonumber\\
\Box t^I&=&(k+3)(k+6), k\ge 0
\eea
Now the cubic action gives the equations of motion
\bea
(\Box- k(k+3))\phi^{I_1}&=&D^{\phi}_{I_1I_2I_3}s^{I_2}s^{I_3}
+E^{\phi}_{I_1I_2I_3}D^{\alpha}s^{I_2}D_{\alpha}s^{I_3}\nonumber\\
&+&F^{\phi}_{I_1I_2I_3} D^{(\alpha}D^{\beta)}s^{I_2}D_{(\alpha}D_{\beta)}
s^{I_3}+...\nonumber\\
(\Box-k_1(k_1-3))s^{I_1}&=&D^{s\phi}_{I_1I_2I_3}\phi^{I_2}s^{I_3}
+E^{s\phi}_{I_1I_2I_3}D^{\alpha}\phi^{I_2}D_{\alpha}s^{I_3}\nonumber\\
&+&F^{s\phi}_{I_1I_2I_3} D^{(\alpha}D^{\beta)}\phi^{I_2}D_{(\alpha}D_{\beta)}
s^{I_3}\nonumber\\&+&D_{I_1I_2I_3}s^{I_2}s^{I_3}
+E_{I_1I_2I_3}D^{\alpha}s^{I_2}D_{\alpha}s^{I_3}\nonumber\\
&+&F_{I_1I_2I_3} D^{(\alpha}D^{\beta)}s^{I_2}D_{(\alpha}D_{\beta)}
s^{I_3}+...
\eea
The condition of getting rid of nonlinear terms with derivatives in 
the equations of motion suggests the rotation:
\bea
\phi^{I_1}&=&\tilde{\phi}^{I_1}+J^{\phi}_{I_1I_2I_3}\tilde{s}^{I_2}\tilde{s}
^{I_3}+L_{I_1I_2I_3}^{\phi}D^{\alpha}\tilde{s}^{I_2}D_{\alpha}\tilde{s}^{I_3}
\nonumber\\
s^{I_1}&=&\tilde{s}^{I_1}+J_{I_1I_2I_3}\tilde{s}^{I_2}\tilde{s}
^{I_3}+L_{I_1I_2I_3}D^{\alpha}\tilde{s}^{I_2}D_{\alpha}\tilde{s}^{I_3}
\nonumber\\
&+&J^{s\phi}_{I_1I_2I_3}\tilde{\phi}^{I_2}\tilde{s}
^{I_3}+L^{s\phi}_{I_1I_2I_3}D^{\alpha}\tilde{\phi}^{I_2}
D_{\alpha}\tilde{s}^{I_3}\label{redef}
\eea
In terms of the redefined fields, the equations of motion become
\bea
(\Box-k(k+3))\tilde{\phi}^{I_1}&=&\lambda^{\phi}_{I_1I_2I_3}\tilde{s}^{I_2}
\tilde{s}^{I_3}\nonumber\\
(\Box-k(k-3))\tilde{s}^{I_1}&=&\lambda^{s\phi}_{I_2,I_1I_3}\tilde{\phi}^{I_2}
\tilde{s}^{I_3}+\lambda_{I_1I_2I_3}\tilde{s}^{I_2}\tilde{s}^{I_3}
\eea
where 
\bea
\lambda^{\phi}_{I_1I_2I_3}&=&-\frac{9k_2!k_3!\Gamma(k_1+5/2)
2^{3k_1-6-3\Sigma/2}}{\Gamma(\alpha_1+1)\Gamma(\alpha_2+1)\Gamma(\alpha_3+1)
\Gamma(\frac{\Sigma _5}{2})}
\frac{(2\alpha_1-3)\Sigma(\Sigma +1)(\Sigma +3)}{k_2(2k_2+1)k_3(2k_3+1)}
\nonumber\\&&
\alpha_1(\alpha_1-1)<T^{I_1}C^{I_2}C^{I_3}>\\
\lambda_{I_1I_2I_3}&=&-\frac{3k_2!k_3!\Gamma(k_1+5/2)2^{3k_1-5-3\Sigma/2}
}{\Gamma(\alpha_1+1)\Gamma(\alpha_2+1)\Gamma(\alpha_3+1) \Gamma(\frac{\Sigma
+5}{2})}\nonumber\\
&&\frac{\Sigma(\Sigma -2)(\Sigma^2-1)(\Sigma^2-9)}{(k_1-1)(2k_1+3)k_2(2k_2
+1)k_3(2k_3+1)}
\alpha_1\alpha_2\alpha_3<C^{I_1}C^{I_2}C^{I_3}>
\eea
and where 
$<C^{I_1}C^{I_2}C^{I_3}>=C^{I_1}_
{i_1...i_{\alpha_2+\alpha_3}}
{C^{I_2\;\; i_1...i_{\alpha_3}}}_{j_1...j_{\alpha_1}
} C^{I_3\;\; i_{\alpha_3+1}...i_{\alpha_3+\alpha_2} j_1...j_{\alpha_1}}$ and 
$<T^{I_1}C^{I_2}C^{I_3}>=T^{I_1\;\;abi_1...i_{\alpha_2+\alpha_3}}
C^{I_2}_{ai_1...i_{\alpha_3}j_a..j_{\alpha_1-1}}{C^{I_3}_{bi_{\alpha_3+1}...
i_{\alpha_3+\alpha_2}}}^{j_1...j_{\alpha_1-1}}$, and $\lambda^{s\phi}$ is
proportional to $\lambda^{\phi}$.
We used the notation:
\bea
\alpha_i&=&\frac{1}{2}\sum_{j=1}^3k_j-k_i\\
\Sigma&=&k_1+k_2+k_3
\eea
The expression $<T^{I_1}C^{I_2}C^{I_3}>$ is non-zero if the 
'modified triangle inequalities' are satisfied:
$\alpha_1\ge 1, \alpha_2\ge 0,\alpha_3\ge 0$, together with the condition 
that $\Sigma $ be even.  

For the s-s-s vertex, if $k_2=k_3=2$, 
we have two possibilities: $k_1=2,4$ . The only massive coupling
($k_1=4$) is extremal, and it vanishes due to the factor of $\alpha_1=0$, 
(this fact also signals the possibility of a consistent truncation to the
 massless sector) but the corresponding CFT correlator is finite, because it is
obtained after multiplying with a factor of
\be 
\frac{\Gamma(2\alpha_1)\Gamma(2\alpha_2)\Gamma(2\alpha_3)}{\Gamma(2k_1-3)
\Gamma(2k_2-3)\Gamma(2k_3-3)}
\ee
and we use that $\Gamma (2\alpha_1)/\Gamma (\alpha_1)\rightarrow 1/2$ when
$\alpha_1\rightarrow 0$. This
analytical continuation procedure was discussed by Liu and Tseytlin
\cite{lt3} who noticed that although the coupling  
dilaton-dilaton-massive singlet ($M^2=32$) vanishes, the 3-point function of
associated CPOs does not. 
For the $\phi -s-s$ vertex,  the rescaling factor is finite, namely
\be
\frac{\Gamma(2\alpha_3+3)\Gamma(2\alpha_1-3)\Gamma(2\alpha_2+3)}{\Gamma(2k_2)
\Gamma(2k_3)\Gamma(2k_1+6)}
\ee
so the CFT correlator computed from $\lambda^{\phi}_{222}$ remains zero.

We will argue that the nonlinear field redefinition (\ref{redef}) 
is not just a 
matter of conveniently getting rid of unwanted higher-derivative terms in the
scalar field action, but it is precisely (when truncated to the massless 
sector) a Taylor expansion of the nonlinear KK ansatz \cite{nvv, nvv2} in the
transverse fluctuation gauge. So, in this respect, it is not an unnatural 
redefinition, but it is the one which gives the gauged sugra action. For
instance a nonlinear redefinition of the 3-index antisymmetric tensor
$a_{\alpha\beta\gamma}\rightarrow a_{\alpha\beta\gamma}+
\epsilon_{ABCDE}F_{[\alpha\beta}^{AB} B_{\gamma ]}^{CD} Y^E+more$ will 
generate part of the 7d CS terms, which we previously argued that are 
absent when one uses the linearized ansatz.
Start with the nonlinear KK metric ansatz
\bea
G_{\alpha\beta}(x,y)&=&\Delta^{-2/5}(x,y)g_{\alpha\beta}(y)\\
G_{\mu\nu}(x,y)&=&\Delta^{4/5}(x,y)C_\mu^A(x) T_{AB}^{-1}(y) C_\nu^B(x)\\
G_{\mu\alpha}&=&2\Delta^{4/5}(x,y) B_\alpha^{AB}(y) Y^B(x) C_\mu^C(x) 
T^{-1}_{AC}(y)
\eea
where $\Delta^{-6/5}=Y\cdot T\cdot Y$, and the spherical harmonic satisfy the 
following identities: $\Box_x Y^A(x)=-4 Y^A(x)$, 
$Y(x)\cdot Y(x)=\sum_{A=1}^5 Y^A Y^A=1$,
$\partial_\mu Y^A(x)=C_\mu^A(x)$ is the conformal Killing vector and
$C_\mu^A {C^\mu}^B=\delta^{AB}-Y^A Y^B$; indices are raised and lowered 
with Kronecker delta.
Set the gauge fields to zero and expand in linear order in the scalar 
fluctuations $\delta\pi_{AB}$. Then
\be
h_{\mu\nu}=2C_\mu\cdot \delta\pi\cdot C_\nu+\frac{4}{3}Y\cdot\delta\pi\cdot Y
\sirc g_{\mu\nu}
\ee 
will not be in the transverse gauge, and we need a compensating Einstein
transformation\footnote{
$\tilde g=\sum\frac{1}{n!} (L_\xi)^n g$ where $L_\xi$ is the Lie derivative 
along $\xi$, and $\tilde g$ is the transformed metric} with parameter 
$\xi_\nu =-C_\nu\cdot\delta\pi\cdot Y$
to satisfy the gauge condition $\sirc D^\mu h_{(\mu\nu)}=0$. Thus,
in the transverse gauge, and up to quadratic order in fluctuations
$h_{(\mu\nu)}=0$. The gauge condition $\sirc D^\mu h_{\mu\alpha}=0$ also
implies the need of another Einstein transformation with parameter
$\xi_\alpha=1/2 Y\cdot \sirc D_\alpha \delta\pi \cdot Y$ and one gets
$h_{\mu\alpha}=0$, after performing the Einstein transformations.
\footnote{
The constraint $(h'^I-9/10 h_2^I)\sirc D_{(\alpha} \sirc D_{\beta )}Y^I=0$
is trivially satisfied on-shell if we restrict ourselves to the 
massless scalar
sector, as it yields the scalar field eq. $Y\cdot (\Box\delta \pi +2\delta\pi)
\cdot Y=0$.} 
To second order in fluctuations, and in the transverse gauge 
we have
\bea
h_{(\mu\nu)}&=&-\frac{4}{9}C_{(\mu}\cdot\delta\pi^2 \cdot C_{\nu )}-
\frac{4}{3}C_{(\mu}\cdot\delta\pi\cdot C_{\nu )} Y\cdot\delta\pi\cdot Y+
\frac{4}{3} C_{(\mu}\cdot\delta\pi \cdot Y C_{\nu )}\cdot \delta\pi\cdot 
Y\nonumber\\
&+&\frac{1}{9} C_{(\mu}\cdot\sirc D_\alpha\delta\pi\cdot\sirc D^{\alpha}\delta
\pi\cdot C_{\nu )}+\frac{1}{3}C_{(\mu}\cdot\sirc D_\alpha\delta\pi\cdot 
C_{\nu )} Y\cdot \sirc D^\alpha \delta\pi \cdot Y\nonumber\\
&-&\frac{1}{3}
C_{(\mu}\cdot\sirc D_\alpha\delta\pi \cdot Y C_{\nu )}\cdot \sirc 
D^\alpha\delta\pi\cdot Y\label{mass}
\eea
where, again we needed a compensating Einstein transformation with 
parameter
\bea
\tilde\xi_\nu&=&
-\frac{20}{9} C_\nu \cdot (\delta\pi)^2\cdot Y-
\frac{1}{12}Y\cdot \sirc D_\alpha \delta\pi\cdot Y
C_\nu\cdot \sirc D^\alpha\delta\pi\cdot Y+\frac{1}{18} Y\cdot\sirc D_\alpha
\delta\pi \cdot\sirc D^\alpha \delta\pi \cdot C_\nu\nonumber\\
\eea
Using now that the first massive mode in $h_{\mu\nu}(y,x)=\phi^I(y) Y_I(x)$
has a spherical harmonic
\bea
Y_{\mu\nu}^{ABCD}&=&C_{(\mu}^{(A} C_{\nu)}^{B)} Y^{(C}Y^{D)}-
\frac{1}{2}C_{(\mu}^{(C}C_{\nu )}^{B)}Y^{(A} Y^{D)}-
\frac{1}{2}C_{(\mu}^{(C}C_{\nu )}^{A)}Y^{(B} Y^{D)}\nonumber\\
&-&\frac{1}{2}C_{(\mu}^{(A}C_{\nu )}^{D)} Y^{(B} 
Y^{C)}-\frac{1}{2}C_{(\mu}^{(B}C_{\nu )}^{D)} Y^{(A} 
Y^{C)}
+C_{(\mu}^{(D} C_{\nu )}^{C)} Y^{(A} Y^{B)}\nonumber\\
&-&\frac{2}{15}\left(\delta^{AB} C_{(\mu}^{(D} C_{\nu )}^{C)}+
\delta^{CD}C_{(\mu}^{(A} C_{\nu)}^{B)}-\delta^{AD}C_{(\mu}^{(C}
C_{\nu )}^{B)}-\delta^{BC} C_{(\mu}^{(A}C_{\nu )}^{D)}\right)
\eea
where the symmetry in the $\{ABCD\}$ indices is given by the box Young 
tableau (therefore it is traceless in any pair of indices) and 
$\Box_x Y_{\mu\nu}^{ABCD}=-8 Y_{\mu\nu}^{ABCD}$, we
notice that (\ref{mass}) can be rewritten as
\bea
h_{(\mu\nu )}&=&\frac{1}{6}(-4\delta\pi^{AB} \delta\pi^{CD}+
\sirc D_\alpha\delta\pi^{AB} \sirc D^\alpha\delta\pi^{CD}) 
Y_{\mu\nu}^{ABCD}
\eea
For the same massive mode the nonlinear redefinition reads:
\be
0=\phi^{ABCD}+const.
\left(-\sirc D_\alpha s^{AB} 
\sirc D^\alpha s^{CD}+4R^{-2} s^{AB} s^{CD}\right)
\ee
where $R$ is the $S_4$ radius (for us, \cite{nvv,nvv2} $R=1$, 
while in \cite{cfm} $R=1/2$).

Thus we explicitly showed that the nonlinear redefinition coincides 
with the nonlinear KK ansatz
in the transverse gauge.
\footnote{
One cannot directly read off from here the relationship between 
$\delta\pi^{AB}$ and $s^{AB}$, one of the reasons being that the 
spherical harmonics were normalized differently in \cite{nvv2} and in
\cite{cfm}.} 

In conclusion, since after the nonlinear rotation we get a 
consistent truncation,
and moreover, we get the correct gauged sugra terms (up to cubic order in 
fluctuations), we can say that the nonlinear 
ansatz in \cite{nvv2, cfm} is the correct one to use in the AdS-CFT 
correspondence.

Let us now discuss the 5-dimensional case. In fact, it is a characteristic
of gauged sugras that the kinetic term is $P_{\alpha ij}P^{\alpha ij}$,
$P_{\alpha ij}=[\Pi^{-1}\partial_{\alpha}\Pi ]_{(ij)}$, with $\Pi$ the scalar 
coset vielbein. It follows that the kinetic term has always two derivatives, 
and we also find no cubic term in $P_{\alpha ij}^2$.
Now, we would like to see that the gauged sugra action is 
the correct one to use for the AdS-CFT correspondence.

Lee et al. \cite{lmrs} looked at the 10d IIB action compactified on 
$AdS_5\times S_5$.
The fluctuations are written as:

\bea
G_{mn}&=&g_{mn}+h_{mn}\nonumber\\
h_{\alpha\beta}&=&h_{(\alpha \beta )}+\frac{h_2}{5}\nonumber\\
h_{\mu\nu}&=&h'_{(\mu\nu)}-\frac{h_2}{3}g_{\mu\nu}+\frac{h'}{5}g_{\mu\nu}
\nonumber\\
F&=&\bar{F} +5\nabla_{[i}a_{jklm]}
\eea

If one decomposes linearly in spherical harmonics as:

\bea
h'_{\mu\nu}&=&\sum h'_{\mu\nu I}Y^I\nonumber\\
h_2&=&\sum h_{2I}Y^I\nonumber\\
a_{\alpha_1\alpha_2\alpha_3\alpha_4}&=&\sum \nabla^{\alpha}\epsilon_{\alpha
\alpha_1\alpha_2\alpha_3\alpha_4}b_IY^I\nonumber\\
a_{\mu_1\mu_2\mu_3\mu_4}&=&\sum a_{\mu_1\mu_2\mu_3\mu_4 I} Y^I
\eea
the constraints on the fields can be solved and the $h_{2}^I,B^I$ system is 
diagonalized  by 
\bea
s^I&=&\frac{1}{20(k+2)}[h_2^I -10(k+4) b^I]\nonumber\\
t^I&=&\frac{1}{20(k+2)}[h_2^I+10kb^I]
\eea
such that 
\bea
\Box s^I&=&k(k-4)s^I,\;\; k\ge 2\nonumber\\
\Box t^I&=& (k+4)(k+8)t^I,\;\; k\ge 0
\eea
If one compactifies the type IIB action in 10d one obtains the following 
equations of motion for the $s$ fields (up to quadratic order in fluctuations)
\bea
(\Box -m_{I_1}^2)s^{I_1}&=& \sum_{I_2,I_3}\left( D_{I_1I_2I_3}
s^{I_2}s^{I_3} + E_{I_1I_2I_3} \nabla_{\mu}s^{I_2}\nabla^{\mu}s^{I_3}
\right.\nonumber\\ 
&&\left. +F_{I_1I_2I_3} \nabla^{(\mu}\nabla^{\nu)}s^{I_2}\nabla_{(\mu}\nabla_{\nu)}
s^{I_3}\right)
\eea
where $D_{I_1I_2I_3}, E_{I_1I_2I_3}$ and $F_{I_1I_2I_3}$ are constants 
depending on $k_1,k_2,k_3$.
But in order to get rid of the terms in the equations of motion 
nonlinear in $s^I$ and involving derivatives, one needs to make a nonlinear 
redefinition of fields,
\bea
s^{I_1}&=&s^{,I_1}+\sum( J_{I_1I_2I_3} s^{,I_2}S^{,I_3} +L_{I_1I_2I_3}
\nabla^{\mu}s^{,I_2}\nabla_{\mu}s^{,I_3}\nonumber\\
J_{I_1I_2I_3}&=&\frac{1}{2}E_{I_1I_2I_3}+\frac{1}{4}F_{I_1I_2I_3}(m_{I_1}^2
-m_{I_2}^2-m_{I_3}^2-8)\nonumber\\
L_{I_1I_2I_3}&=&\frac{1}{2}F_{I_1I_2I_3}
\eea
which modifies the equations of motion to:
\be
(\Box -m_{I_1}^2)s^{I_1} =\sum _{I_2, I_3}\lambda_{I_1 I_2I_3} s^{I_2}
s^{I_3}
\ee
where
\be
\lambda_{I_1I_2I_3}=\frac{k_1!k_2!k_3}{\alpha_1!\alpha_2!\alpha_3!}\frac{(k+1)
2^{k+2-\Sigma/2}\Sigma ((\Sigma/2)^2-1)((\Sigma/2)^2-4)}{k_1(k_1-1)
(k_2+1)(k_3+1)(\Sigma/2+2)!}\alpha_1\alpha_2\alpha_3<C^{I_1}C^{I_2}C^{I_3}>
\ee
We notice that the invariant tensor $<C^{I_1}C^{I_2}C^{I_3}>$ is nonzero 
(one can contract the indices correctly) only if the 'triangle inequalities'
$\alpha_i\ge 0$ are satisfied, together with the the condition that $\Sigma$
is even. If the coupling $\lambda_{I_1I_2I_3}$ corresponds to a 
massive mode and two massless modes, i.e. $k_2=k_3=2$, then $k_1$ is 
restricted to be 2 (massless) or 4 (massive). The latter case is 'extremal', 
in the sense that $\alpha_1$ takes the extreme value zero).
But in the extremal case $\lambda_{I_1I_2I_3}=0$ because the 
$\alpha_1$ factor vanishes.

The fact that after the  nonlinear redefinition one has a consistent 
truncation, both in 5 and in 7 dimensions, was noticed already in a paper
we wrote with Peter van Nieuwenhuizen \cite{nvv}, but we gave no 
details there. 
Afterwards, Aryutunov and Frolov wrote a series of papers where they 
found similar results. In \cite{arufro,arufro1}, the cubic and quartic terms 
in the action were calculated by expanding the 10d IIB action in scalar 
fluctuations, and after a nonlinear redefinition of fields
they also found a consistent truncation to the massless sector
(the calculation also involves more fields, not just the $s^I$ scalars), 
i.e. all couplings between massless scalars and one massive scalar vanish. 

Moreover, in \cite{arufro2} they find that the corresponding action
for the massless scalars (to quartic order) coincides with the terms 
in the gauged sugra action. 

After finding the coupling $\lambda_{I_1I_2I_3}$, one can use it to compute 
correlators of CPO's in the boundary CFT  using the formulas in \cite{fmmr} 
which are found to agree with the weak coupling 
result. But for that one needed a nonlinear redefinition of fields, that is 
effectively modifying the linear ansatz in \cite{krn} to a nonlinear one. 
This nonlinear
ansatz gives a consistent truncation, because one can consistently put all the
massive $s^I$ in \cite{krn} to zero. We notice that when going from 
a 3-point coupling
in AdS space to a correlator on the boundary, one picks up a factor
\be
\frac{\Gamma(\alpha_1)\Gamma(\alpha_2)\Gamma(\alpha_3)}{\Gamma(k_1-2)
\Gamma(k_2-2)\Gamma(k_3-2)}
\ee
The fact that the denominator becomes infinite is absorbed in the normalization
of the operators.\footnote{ 
The 2-point function of CPOs coupled to the  $k=2$ scalar
fields behaves $\sim (k-2)^2$, and therefore vanishes. To get a nonvanishing 
result (in accord with the CFT calculation) one has to rescale the 
supergravity fields with the infinite factor $1/(k-2)$. This can be 
interpreted as another example of analytical continuation for the `extremal' 
correlator $k_1=k_2=2$.} 
We notice that the $\Gamma(\alpha_1)$ in 
the numerator becomes infinite, so that the 'extremal' correlator becomes
nonzero. We will say more about that at the end of this section.

In the $AdS_5\times S_5$ case, no fully consistent KK truncation is known, but 
it is generally believed that one exists. If this is the case, the procedure 
of taking the consistent truncation is seen to be the correct one. 

Another point to be stressed is that before the rotation, the action has a term
cubic in scalars, but with two derivatives. We have seen that in the gauged 
sugra action we don't have such a term, so the fact that the nonlinear rotation
removes it is another confirmation of our procedure.

Finally, we shall comment on the calculation in \cite{dfmmr}. This paper
tried to address the following puzzle raised by the calculation in 
\cite{lmrs}. If one takes the limit when $k_1\rightarrow k_2 +k_3$ in 
the calculation of \cite{lmrs}, the coefficient of the cubic action for 
the scalars tend to zero, but the integration diverges in such a way that 
the 3-point function becomes zero. 

To address this issue, D'Hoker et al. study the $t\phi\phi$ three point 
function, and instead of using the nonlinear redefinition of fields 
used in \cite{lmrs} (as we argue that is the correct procedure), 
use equations of motion and partial integration to 
arrive at
\bea
&&2k_5^2 S_{cubic}=-8\frac{(\Sigma +4)\alpha_1(\alpha_2 +2)(\alpha_3 +2) }
{(k_1 +3)}\int _{AdS_5} a(k_1,k_2,k_3)t^{k_1}\phi^{k_2}\phi^{k_3}+\nonumber\\
&&\int_{\partial (AdS_5)} \frac{a(k_1, k_2, k_3)}{k_1+3}(-D_n\phi^{k_3}
D_{\mu}\phi^{k_1} D^{\mu}\phi^{k_2} -D_n\phi^{k_2}D_{\mu}t^{k_1}D^{\mu}
\phi^{k_3} \nonumber\\&&
+D_n t^{k_1}D_{mu}\phi^{k_2}D^{\mu}\phi^{k_3})
+{\rm contact\;\; terms}\label{freedman}
\eea
where $\Sigma =\frac{1}{2}(k_1 +k_2 +k_3), \alpha_1=\frac{1}{2}(k_2 +k_3 -
k_1)$, etc., obtaining what we described at the beginning of this section,
namely that the difference between making a nonlinear redefinition of fields 
and using equations of motion and partial integrations is given by
 boundary terms. 
The boundary terms of the type in  (\ref{boundary}) are contact terms which 
were dropped, and the boundary terms in (\ref{boundary2}) are of the same 
type as the ones in (\ref{freedman}). We indeed notice that the coefficient 
of the bulk integral in (\ref{freedman}) becomes equal to zero for $k_1
=k_2+k_3$. The point of view adopted in \cite{dfmmr} is the following. At 
$k_1 < k_2 +k_3$ only the bulk integral contributes, and the boundary one 
doesn't. But at $k_1 =k_2 +k_3$, the situation is reversed: only the boundary
integral contributes, and the boundary one doesn't. Moreover, the result
for $k_1=k_2+k_3$ coincides with the one from the limit $k_1\rightarrow 
k_2 +k_3$. 

Our point of view is that we need to start with only the bulk integral in 
(\ref{freedman}) (in other words make the nonlinear redefinition of fields).
The analytic continuation 
$k_1\rightarrow k_2 +k_3$  gives the correct result. With the linearized 
ansatz one also gets the boundary  integral. If one considers it to be nonzero
as \cite{dfmmr} does, then one can only spoil the result by a factor of 2, 
in this example. 

We note that for this case of scalar fields, this boundary 
terms seem to contribute only for 'extremal correlators' ($k_1=k_2 +
k_3$), with a singular limit needed to be taken, but for general 
fields (gauge fields, for instance) the same will probably not happen.

Indeed, as an example, for gauge fields we saw that the Chern-Simons term is 
completely 
missed by the linearized ansatz. A nonlinear redefinition in 7d which would 
give it
would have to involve $\epsilon_{\alpha_1...\alpha_7}$. And for such a 
redefinition the $A_{\mu}$ equation of motion ($(\Box\delta_{\mu\nu}
-\partial_{\mu}\partial_{\nu})A_{\mu}=0$ ) is not very useful either in terms
of creating the wanted Chern-Simons term by partial integration and use 
of the equations of motion. Neither the natural redefinition one would think 
of, $A_{\alpha_1}\rightarrow A_{\alpha_1}+\epsilon_{\alpha_1...\alpha_7}
\partial^{\alpha_2}A^{\alpha_3}\partial^{\alpha_4}A^{\alpha_5}\partial^
{\alpha_6}A^{\alpha_7}$, nor any other combination is able to reproduce the 
required Chern-Simons term. To be explicit, the nonlinear redefinition
has to involve massive fields being redefined too: massive $\rightarrow$
massive +(massless)$^n$. Similarly, for the use of equations of motion
one needs to use the {\em massive} equations of motion to obtain a 
Chern-Simons term. 

But if one takes as a starting point the linearized ansatz, one will obtain
no anomalous CFT correlators. That is because for that one needs an $\epsilon
$ symbol. In 5 dimensions, the anomaly should be in the 3 point function
already, but clearly the 3 point vertex is non-anomalous (because the $
\epsilon$ symbol comes only from the Chern-Simons which is now absent). In 
7 dimensions, the anomaly starts at the 4 point function. So even though 
there is no Chern-Simons term, one might hope that the exchange diagram,
where massive fields are exchanged, could contribute. But using the 
linearized ansatz we don't get any couplings involving the $\epsilon$ 
symbol of the type gauge field-gauge field-massive field. Therefore again, no 
anomaly on the CFT side. From this discussion, one concludes that, even 
if one does the steps in \cite{dfmmr}, namely partial integration and the 
use of the bulk equations of motion, if one obtains the Chern-Simons term,
it will be together with extra boundary terms canceling the effect of the 
anomaly in the boundary correlators! 

So we have an example where, even if the methods of \cite{dfmmr} can be 
applied, the boundary terms which are generated will contribute not only 
to 'extremal' correlators, but to the 'massless' ones as well.

Therefore, here (for the scalar field case) it is somewhat a matter of taste 
which philosophy one takes, maybe
the one of D'Hoker et al. looks more attractive, however one has to consider 
a more general case. As we have seen, we have an argument that taking the 
nonlinear ansatz from the start produces the correct anomaly, the correct 
gauge invariant AdS correlators (so correct R-invariant CFT correlators) and 
gets rid of unwanted contact terms. Therefore, it is better to have a clear 
physical principle to deal with all of the above. 

For the 'massless' (gauged sugra) AdS fields, we know that gauge invariance
and susy forces us to take the gauged sugra action. For the massive fields, 
we don't know the nonlinear ansatz. But then we can do what Seiberg et al. 
and Corrado et al. did, namely to find order by order and ansatz which removes 
unwanted terms in the action (with too many derivatives, for instance). 
Ideally, one would have to do what we sketched in the introduction, namely 
to use gauge invariance and susy to fix the nonlinear ansatz for the tower
of massive fields.

\section{Conclusions}

In this paper, we gave arguments for the
 previous conjecture that one has to use 
the gauged sugra action, as obtained from 11d, or 10d IIB sugra through a 
nonlinear KK ansatz, for the computation of n-point functions of CFT 
operators coupled 
to the massless (sugra) sector. A similar nonlinear rotation is needed for 
the massive tower in order to restore the gauge invariance of the action for 
the whole KK tower after performing the ``nonlinear rotation'' at the 
massless level.
Part of this rotation, namely up to quadratic order in scalar fluctuations was
 already introduced by Lee et al. \cite{lmrs}, Corrado et al. \cite{cfm} and
Bastianelli and Zucchini \cite{bz}. But their reason for doing this was to 
eliminate certain higher derivative couplings from the reduced action.

Our arguments are based on:\\
- previously computed R-current correlators. In particular, we noticed that 
the linear KK ansatz completely misses the CS term 
(in both 5 and 7 dimensions) 
which corresponds to the R-current anomaly.\\
-we explicitly showed that the nonlinear rotation of \cite{cfm, bz} 
corresponds to a Taylor expansion of the nonlinear KK ansatz \cite{nvv, nvv2} 
(in the transverse gauge) in massless scalar fluctuations.
  
{\bf Acknowledgements} We would like to thank Peter van Nieuwenhuizen, Fiorenzo
Bastianelli, Iosif Bena, Calin Lazaroiu, Radu Roiban and Arkady Tseytlin 
for valuable discusions.

\end{document}